\def \SAIT #1 #2 {{\em Mem.\ Soc.\ Astron.\ It.\/} {\bf #1}, #2}
\def \MESS #1 #2 {{\em The Messenger\/} {\bf #1}, #2}
\def \ASTRNACH #1 #2 {{\em Astron. Nach.\/} {\bf #1}, #2}
\def \AAP #1 #2 {{\em Astron. Astrophys.\/} {\bf #1}, #2}
\def \AAL #1 #2 {{\em Astron. Astrophys. Lett.\/} {\bf #1}, L#2}
\def \AAR #1 #2 {{\em Astron. Astrophys. Rev.\/} {\bf #1}, #2}
\def \AAS #1 #2 {{\em Astron. Astrophys. Suppl. Ser.\/} {\bf #1}, #2}
\def \AJ #1 #2 {{\em Astron. J.\/} {\bf #1}, #2}
\def \ANNREV #1 #2 {{\em Ann. Rev. Astron. Astrophys.\/} {\bf #1}, #2}
\def \APJ #1 #2 {{\em Astrophys. J.\/} {\bf #1}, #2}
\def \APJL #1 #2 {{\em Astrophys. J. Lett.\/} {\bf #1}, L#2}
\def \APJS #1 #2 {{\em Astrophys. J. Suppl.\/} {\bf #1}, #2}
\def \APSS #1 #2 {{\em Astrophys. Space Sci.\/} {\bf #1}, #2}
\def \ASR #1 #2 {{\em Adv. Space Res.\/} {\bf #1}, #2}
\def \BAIC #1 #2 {{\em Bull. Astron. Inst. Czechosl.\/} {\bf #1}, #2}
\def \JSQRT #1 #2 {{\em J. Quant. Spectrosc. Radiat. Transfer\/} {\bf
#1}, #2}
\def \MN #1 #2 {{\em Mon. Not. R. Astr. Soc.\/} {\bf #1}, #2}
\def \MEM #1 #2 {{\em Mem. R. Astr. Soc.\/} {\bf #1}, #2}
\def \PLR #1 #2 {{\em Phys. Lett. Rev.\/} {\bf #1}, #2}
\def \PASJ #1 #2 {{\em Publ. Astron. Soc. Japan\/} {\bf #1}, #2}
\def \PASP #1 #2 {{\em Publ. Astr. Soc. Pacific\/} {\bf #1}, #2}
\def \NAT #1 #2 {{\em Nature\/} {\bf #1}, #2}

\documentstyle{memsait}
\input epsf.sty
\begin{opening}
\title{SPECTRAL VARIABILITY OF QSOs IN THE OPTICAL BAND}
\author{DARIO TR\`EVESE$^1$, FAUSTO VAGNETTI$^2$}
\institute{$^1$Dipartimento di Fisica, Universit\`a ``La Sapienza'',
Roma, Italy\\
$^2$Dipartimento di Fisica, Universit\`a ``Tor Vergata'', Roma, Italy}
\date{} % DO NOT INSERT ANY DATE HERE !!!
\end{opening}

\begin{document}

%\oddpagefooter{\sf Mem. S.A.It., Vol. ??, ??}{}{\thepage}
%\evenpagefooter{\thepage}{}{\sf Mem. S.A.It., Vol. ??, ??}
\oddpagefooter{}{}{} % LEAVE AS IT IS !
\evenpagefooter{}{}{} % LEAVE AS IT IS !
\
\bigskip

\begin{abstract}
A new analysis of the variability of the spectral slope of PG QSOs has
been performed, on the basis of recent literature data in the B and R
photometric bands.  Preliminary results confirm our previous findings
concerning the increase of variability with the rest-frame observing
frequency.  We also find a correlation of the spectral slope with
luminosity, consistent with temperature changes of an emitting black
body.
\end{abstract}
\section{Introduction}
Although variability plays a key role in constraining the models of
the central engine of QSOs, little is known about the physical origin
of luminosity variations.  The most diverse variability mechanisms
have been proposed in the past, including supernovae explosions
(Aretxaga et al. 1997), instabilities in the accretion disk (Kawaguchi
et al. 1998), and gravitational lensing due to intervening matter
(Hawkins 1993). So far, most of the information about the
characteristics of variability derives from the statistical analysis
of single band light curves of magnitude limited QSO samples. The
correlation of variability with either intrinsic luminosity ($v$-$L$)
or redshift ($v$-$z$) is affected by the strong correlation between
luminosity and redshift ($L$-$z$), present in these samples. The
results of these analyses also depend on the specific variability
index adopted, as shown by Giallongo et al. (1991, GTV), who found a
positive $v$-$z$ correlation through a variability index based on the
rest-frame structure function, later confirmed by Cristiani et
al. (1996). The increase of variability with the rest-frame observing
frequency found by Di Clemente et al. (1996), supports the suggestion
of GTV that QSOs at high redshift appear more variable, on average,
since they are observed in a higher rest frame frequency. The
dependence of variability on frequency is associated with the
hardening of the spectral energy distribution (SED) during the bright
phases (Cutri et al. 1985, Kinney et al. 1991, Edelson et al. 1990).
Tr\`evese et al. (1999) have shown that a hardening of the SED in the
bright phase occurs, on average, in the statistical, magnitude
limited, sample of AGNs of the SA 57 (Tr\`evese et al. 1989, 1994,
Bershady et al. 1998). They also found that the slope $\alpha$ and its
variations associated with the luminosity changes, are consistent on
average with temperature changes of a black body.
\section{Analysis and results}
In a recent paper, Giveon et al. (1999) provide the light curves of a
subsample of PG quasars consisting of 42 nearby and bright QSOs
($z<0.4$, $B<16$ mag) which were monitored for 7 years in the Johnson
Cousins B and R bands, with the 1 m telescope of the Wise Observatory,
with a typical sampling interval of 39 days.  We define the structure
function as (see Di Clemente et al. 1996)
$S=\sqrt{\pi/2 \langle |m(t+\tau)-m(t)|\rangle^2-\sigma_n^2}$,
where $m(t)$ is either the B or the R magnitude, t is the rest-frame
time, $\tau$ is the time lag between the the observations, $\sigma_n$
is the relevant r.m.s. noise and the angular brackets indicate the
average taken over all the pairs of observations separated by a time
interval $\tau \pm \Delta\tau$.  The value of $\Delta\tau$ is the
result of a trade-off between time resolution and statistical
uncertainty.  In the following we define, for each QSO and for each
band, two different variability indexes $v_1$ and $v_2$ defined as
$S$, computed for $\tau=0.3\pm 0.09$ yr and for $\tau=2\pm 0.6$ yr
respectively.  The average values of $v_{1B}$, $v_{1R}$, $v_{2B}$,
$v_{2R}$ over the ensemble of 42 QSOs, versus the relevant average
rest-frame frequency, are reported in Figure 1a, which is adapted from
Di Clemente et al. (1996).
\begin{figure}
\epsfxsize=10cm % fix the x-dimension and scales y-dim. to x-dim.
% Feel free to do the choice you prefer but do not exceed the
% x-dimension of the text lines. for centering: act on hspace argument.
~\hspace{1.5cm}\epsfbox{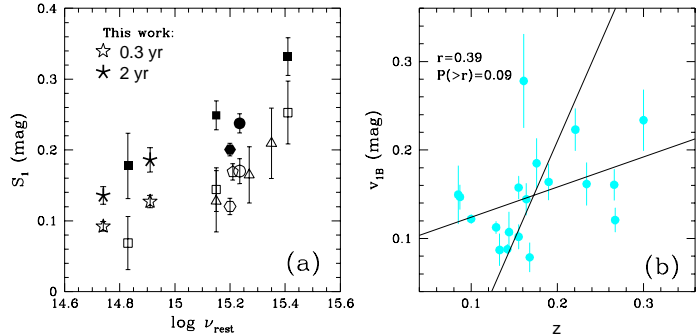}
\vspace*{-5.4cm}
\caption[h]{(a) Variability vs. rest-frame frequency for various QSO
samples, adapted from Di Clemente et al. 1996. New points are represented by
stars.
(b) Variability vs. redshift for the subsample with $-23.5<M_B<-22.5$.}
\end{figure}
The new points are consistent with the general trend and confirm the
value of the slope $\partial S_1 / \partial \log \nu_{rest} = \partial
S_1 / \partial \log (1+z) \simeq 0.25 - 0.3$ which accounts for the
$v$-$z$ correlation found by GTV.  Although this correlation is not
present for the whole sample (Giveon et al. 1999), due to the small
redshift range ($z<0.4$), the $v$-$z$ correlation appears (see
Fig. 1b), if we restrict to an absolute magnitude bin $-23.5 < M_B <
-22.5$, to reduce the effect of the (positive) $L$-$z$ and (negative)
$v$-$L$ correlations. The correlation coefficient $r_{v,z} = 0.39$ is
only marginally significant ($P(>r)=0.09$), since the number of QSOs
in the selected magnitude bin is only 19. Therefore, the $v$-$z$
correlation and its interpretation in terms of spectral variability
are confirmed, despite the poor statistics and the narrow redshift
range. For each QSO we computed the instantaneous spectral slope
$\alpha(t) \equiv \log (f_{\nu_B}/f_{\nu_R})/\log(\nu_B/\nu_R)$,
regarding as synchronous the observations within a time lag of 9
hours.  In Figure 2 $\alpha(t)$ is shown for each QSO at each
observing time as a function of the monochromatic luminosity $L_\nu$
at a fixed rest-frame frequency.  The regression lines $\alpha$
vs. $\log L_{\nu}$ are also reported for each QSO.  With a few
exceptions, a general trend of increasing $\alpha$ for increasing
$\log L_{\nu}$ appears.  Moreover, the average $\alpha$ of each QSO is
also positively correlated with the average luminosity. This fact
suggests that both the increase of $\alpha$ during luminosity
variations and the average increase of $\alpha$ from faint to bright
QSO have similar physical origins, like e.g. the increase of
temperature of the emitting gas.  For comparison in Figure 2 are
reported the $\alpha(t)$ versus $\log L_{\nu}(t)$ lines computed for
three emitting black bodies of different area.
\begin{figure}
\epsfxsize=9cm % fix the x-dimension and scales y-dim. to x-dim.
% Feel free to do the choice you prefer but do not exceed the
% x-dimension of the text lines. for centering: act on hspace argument.
~\hspace{2.2cm}\epsfbox{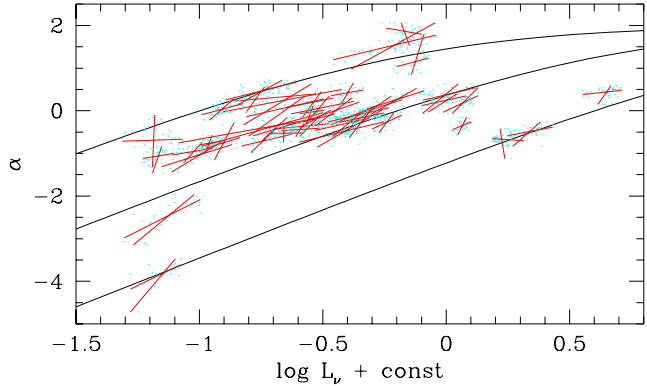}
\vspace*{-3.6cm}
\caption[h]{Spectral slope $\alpha$ vs. monochromatic luminosity
at each observing time. Regression lines $\alpha$-$L_\nu$ are reported
for each QSO.  The three curves represent black bodies of different
areas, with temperature as a free parameter.}
\end{figure}
This admittedly oversimplified model accounts qualitatively for both
the intra-QSO and the inter-QSO $\alpha(t) - \log L_{\nu}(t)$
correlation.  Future comparison with different emission models will
indicate their consistency or inconsistency with SED variability data.

% References. We avoided using the \bibitem command since we found it is
% somewhat platform-dependent. We also avoided using the \cite{keyword}
% command since we found it cumbersome. However, if you are an expert
% LateX user you may use the various LateX tools for the references
% provided they give the same printout formats of the examples given here.

\end{document}